# Non-dipole effects in two-photon double ionization of the *K*-shell of a beryllium-like atomic ion

Alexey N. Hopersky, Alexey M. Nadolinsky, Rustam V. Koneev, and Julia N. Kolesnikova

*Rostov State Transport University, 344038, Rostov-on-Don, Russia*

*E-mail*: qedhop@mail.ru, amnrnd@mail.ru, koneev@gmail.com, juli_tn@mail.ru

**Abstract.** In the second order of the non-relativistic quantum perturbation theory and outside the framework of the dipole approximation for the operator of the radiation transition between continuum-spectrum states, the analytical structures and absolute values of the generalized cross-sections of the two-photon double ionization of the *K*-shell of beryllium-like ions of titanium ($Ti^{18+}$), iron ($Fe^{22+}$) and zinc ($Zn^{26+}$) atoms were predicted. It has been established that taking into account non-dipole effects by several orders of magnitude (*giant* non-dipole effect) reduces the generalized cross-sections calculated within the framework of the dipole approximation. It has also been established that at high (12.5 – 28 keV) energies of the absorbed photons, the generalized cross-section of the two-photon double ionization of the *K*-shell is several orders of magnitude greater than the generalized cross-section of the single ionization. At the same time, as was to be expected, the transition from a neon-like ion to a beryllium-like ion is accompanied by a significant increase in the role of non-dipole effects.

---

## Introduction

The effect of the production of states with an "empty" *K*-shell of an atom [1] is experimentally recorded, in particular, by observing $K^{h}_{\alpha,\beta}$ - hypersatellites of X-ray emission spectra and a fine Auger– and photoelectron- resonance structure spectra when an atom absorbs synchrotron radiation [2,3] and X-ray free-electron laser (XFEL) radiation [4,5]. Experimental studies of the generalized cross-section of the two-photon double ionization of the *K*-shell when absorbed by an atom (atomic ion), for example, XFEL- radiation of high peak brightness and ultra-short duration of the photon pulse [6,7] are still lacking. The first theoretical study of the generalized cross-section of the direct (through the production of the intermediate $1sxp\,(^1P_1)$ state of single ionization of the atom) of the two-photon double ionization of the *K*-shell on the example of the neon atom within the framework of the dipole approximation for the operator of the radiation transition between continuum-spectrum states was carried out in [8]. In the paper [9] goes beyond the dipole approximation and discovers a *gigantic* non-dipole effect on the example of two-photon double ionization of the *K*-shell of a neon-like ion of an iron atom ($Fe^{16+}$). In this article, we generalize the theory of work [9] on the sequence of beryllium-like atomic ions. The purpose of this generalization is to study the dynamics of non-dipole effects with the disappearance of the outer $2p^6$ - shell during the transition from a neon-like ion to a beryllium-like ion. The objects of study are beryllium-like ions of titanium ($Ti^{18+}$; the charge of the nucleus of the ion $Z$ = 22, the configuration and term of the ground state $[0]=1s^2 2s^2\,(^1S_0)$), iron ($Fe^{22+}$; $Z$ = 26) and zinc ($Zn^{26+}$; $Z$ = 30) atoms. The choice is due to the spherical symmetry of the ground states of ions and their availability in the gas phase [10] during the XFEL - experiment [11].

## 1. Theory

Consider the following two-photon double ionization channels of the *K*-shell of a beryllium-like atomic ion:

$$2\hbar\omega+[0]\rightarrow 1sxp+\hbar\omega\rightarrow 1s^0 yl_1 zl_2\,, \qquad (1)$$

$$2\hbar\omega\geq I(1s^{-2})\,. \qquad (2)$$

In (1), (2) the $2s^2$ – shell is not specified, $x, y, z$ - the energies of electrons of the continuous spectrum, $l_1l_2 = ss\,(^1S_0),\, sd\,(^1D_2)$, $\hbar$ - the Planck's constant, $\omega$ - the circular frequency of the absorbed photon, and $I(1s^{-2})$ are the energy of the threshold for the formation of the "empty" *K*-shell. In qualitative agreement with the work [9], the contribution of the channel $l_1l_2 = pp\,(^1S_0,^1D_2)$ to the generalized cross-section turned out to be several orders of magnitude less than the contributions of the channels $l_1l_2 = ss$ , *sd* and in (1) was not taken into account. The contribution of the channel $2\hbar\omega + [0] \to 1sxl \to 1s^0 yl_1 zl_2$, $l \in [0;\infty)$ to the generalized cross-section is determined by the *contact* transition operator and is proportional to square of the matrix element $\langle 1s\,|\,j_l\,|\,xl\rangle$, where $j_l$ is the spherical Bessel function. For a matrix element with a localized wave function $1s$–electron at the energies of the absorbed photons studied in this article, the condition for the applicability of the *dipole* approximation is satisfied: $\lambda_\omega / r_{1s} >> 1$. Here $\lambda_\omega$ is the wavelength of the absorbed photon and $r_{1s}$ is the average radius of the $1s^2$ - shell. As a result, $j_0 \to 1$, $j_{l\geq 1} \to 0$ and $\langle 1s\,|\,j_l\,|\,xl\rangle \to 0$ (see [12] for ion $Ti^{18+}$). The requirement to meet inequality (2) corresponds to the exclusion in (1) of the excited (both intermediate and final) states of the discrete spectrum. The strong spatial and energetic separation of the valence $2s^2$ - shell from the deep $1s^2$ - shell of the ion (see [12] for ion $Ti^{18+}$) makes it possible to neglect the production of finite $(1s2s, 2s^0)yl_1zl_2$ ionization states.

The probability amplitudes of the two-photon double ionization of the *K*-shell ion by channels (1) are physically interpreted in Fig. 1 in the Feynman diagram formalism.

The following formulas of this section of the article are given in atomic units ($e = \hbar = m_e = 1$). In the following sections of the article, ordinary units are restored.

Within the framework of the accepted approximations, the analytical structures for the *partial* generalized cross-sections of the two-photon double ionization of the *K*-shell of the ion through channels (1) are as follows [9]:

$$\sigma^{(ss)} = \mu K^2 \cdot J\,, \tag{3}$$

$$\sigma^{(sd)} = \frac{6}{5}\left(1 - \frac{1}{4\pi}\right) \cdot \sigma^{(ss)}, \tag{4}$$

$$K \cong (\omega\sqrt{2\varepsilon} / \gamma_{1s}^2) \cdot \langle 1s_0\,|\,\hat{r}\,|\,\varepsilon p_+\rangle, \tag{5}$$

$$J = \frac{a}{3}\left\{8 f^2(a/2) + [f(0) + f(a)]^2)\right\}. \tag{6}$$

Here the values are defined: $\mu = 0.278\cdot 10^{-52}$ [cm$^4\cdot$s], $\varepsilon = \omega - I_{1s}$, $I_{1s}$ – the energy of the ionization threshold of the $1s^2$ – shell, $\gamma_{1s} = \Gamma_{1s}/2$, $\Gamma_{1s}$ – the width of the breakdown of the $1s$ – vacancies, overlap integral $f(y) = \langle 1s_+\,|\,ys_{++}\rangle$, $a = 2\omega - I(1s^{-2})$ and the amplitude of $1s \to \varepsilon p$ ionization probability (Fig. 1 at $t = t_1$ and $x \to \varepsilon$):

$$\langle 1s_0\,|\,\hat{r}\,|\,\varepsilon p_+\rangle = \int_0^\infty P_{1s_0}(r) \cdot r \cdot P_{\varepsilon p_+}(r)\,dr\ . \tag{7}$$

The indices "0, +, ++" in (5), (6) and (7) correspond to $P(r)$–the radial portions of the wave functions of electrons obtained by solving the self-consistent Hartree-Fock field equations for [0] –, $1s_+\varepsilon p_+$ – and $1s^0 ys_{++}z(s,d)_{++}$ – ion configurations.

The function $K$ of (5) is obtained in *dipole* approximation for the complete radiative transition operator:

$$\hat{R} = -\frac{1}{c}\sum_{n=1}^{N}(\hat{p}_n \cdot \hat{A}_n)\,. \tag{8}$$

In (8) the following are determined: $c$ – the speed of light in a vacuum, $\hat{p}_n$ – the momentum operator of the $n$-electron of the ion, $\hat{A}_n$ – the operator of the electromagnetic field (in the representation of secondary quantization) and $N$ – the number of electrons in the ion. For the matrix element $\langle xp_+ | \hat{r} | z(s,d)_{++} \rangle$, $r \in [0;\infty)$ (Fig. 1 for $t = t_2$) transition between continuous-spectrum states, the dipole approximation loses its physical meaning (due to the *non-localizability* of the continuous-spectrum), and it is necessary to take into account non-dipole effects. Such accounting through the *integral Loudon representation* for the non-relativistic $\hat{R}$ - operator [13] is carried out in the works [14, 15]. In these works, it was shown that taking into account the decomposition of the operator $\hat{A}_n$ into an infinite functional series by multipoles [16] is equivalent to a simple approximation of the substitution for a single-electron operator of a radiation transition:

$$\hat{r},\ r \in [0;\infty) \to D_\omega = \begin{cases} \hat{r},\ r \in [0;\ q), \\ q, r \in [q;\ \infty). \end{cases} \tag{9}$$

In (9), the magnitude $q = (3/8)\lambda_\omega$ is determined by the behaviour of the $j_1$–spherical Bessel function (at $r \to \infty$, oscillating, $j_1 \to 0$), depends only on the energy of the absorbed photon and turns out to be the "*stopping point*" of the infinite growth of the operator $\hat{r}$. Then, taking into account (9) and the inequality $\gamma_{1s} << 4\varepsilon$, the function $K$ of (5) is modified by substituting:

$$K \to K \cdot \Psi(\omega), \tag{10}$$

$$\Psi(\omega) = 1 - \exp\{-(\gamma_{1s} / \sqrt{2\varepsilon}) \cdot q\}. \tag{11}$$

It should be noted that the analytical results of this section of the article were obtained under the assumption of conducting an XFEL - experiment with *linearly* polarized photons.

## 2. Results and discussion

The calculation results are shown in Fig. 2. The values of the parameters of the generalized cross-sections are given in Table 1.

The transition from a neon-like ion to a beryllium-like ion is accompanied by a significant decrease in the parameter $\Gamma_{1s}$. For example, $\Gamma_{1s}$($Fe^{16+}$) = 1.0460 eV [17] >> $\Gamma_{1s}$($Fe^{22+}$) = 0.0909 eV [18]. In fact, the "loss" of the $2p^6$-shell of the ion "removes" the leading contribution of the radiation-free Auger - decay channel $1s(2s^2)2p^6 \to 1s^2(2s^2)2p^4xl$ to the total width of the decay of the 1*s*- vacancy. Here $x$ is the energy of the Auger - electron of the continuous spectrum, $l = s, d$ and the $2s^2$-shell plays the role of an "observer". As a result, in the studied regions of the energies of the absorbed photons, the values of the $\Psi(\omega)$–function (11) for the beryllium-like ion decreases, and Fig. 2 shows a much more pronounced gigantic non-dipole effect than for the neon-like ion:

$$\eta = \sigma_D / \sigma_{ND} \cong 10^8 >> 10^6, \tag{12}$$

$$\sigma = \sigma^{(ss)} + \sigma^{(sd)}. \tag{13}$$

In (12) are determined: the indices *D* (*ND)* – dipole (non-dipole) approximation for the radiation transition operator and $\eta \cong 10^6$ - see Fig. 3 of the paper [9] for a neon-like ion. It should be noted here that the concept of a "*giant non-dipole effect*" also includes the following fact. The transformation of the *K*-function (10) gives not only the result (12), but also leads (formally mathematically $\Psi(\omega)$ ) $\to 0$ at $\hbar\omega \to \infty$) to a significant "*localization"* of the energy region of the absorbed photon, where the effect of two-photon double ionization of the *K*-shell *still occurs* (see Figs. 2b, c of the paper [9]).

Along with the result (12), the "loss" (shielding the "empty" *K*-shell of the ion) of the $2p^6$-shell leads to a significant increase in the ratio of the generalized cross-sections of the two-photon double and single ionization of the *K*-shell of the beryllium-like ion:

$$\zeta = (\sigma_{ND}^{(ss)} + \sigma_{ND}^{(sd)})/(\sigma^{(s)} + \sigma^{(d)}) \cong 10^{12} >> 10^3. \quad (14)$$

In (14) the sum $(\sigma^{(s)} + \sigma^{(d)})$ is taken from [12] and $\zeta \cong 10^3$ – see formula (52) of work [9] for a neon-like ion. The results (14) can be given the following physical interpretation. The probability of a single ionization is determined by the amplitudes of radiation $1s \to \bar{\varepsilon}p$ and $np \to \bar{\varepsilon}(s,d)$ transitions with $\bar{\varepsilon} = 2\omega - I_{1s}$ (see formulas (22) ÷ (26) of the work [12]). The probability of double ionization, according to (5), is determined by the amplitude of the radiation $1s \to \varepsilon p$ transition with $\varepsilon = \omega - I_{1s}$. Then, by virtue of $\bar{\varepsilon} >> \varepsilon$, we have "strong" inequalities: $\left|\langle 1s_0 | \hat{r} | \bar{\varepsilon}p_+ \rangle\right|$, $|\langle np_+ | \hat{r} | \bar{\varepsilon}(s,d)_+ \rangle|$ << $\left|\langle 1s_0 | \hat{r} | \varepsilon p_+ \rangle\right|$. At the same time, the contribution of the product of amplitudes $\langle 1s_0 | \hat{r} | np_+ \rangle \langle np_+ | \hat{r} | \bar{\varepsilon}(s,d)_+ \rangle$ is noticeably suppressed in the region of the energies of the absorbed photon $\hbar\omega \geq I_{1s}$ by the energy denominator with resonant energy $\hbar\omega = I_{1snp}$, where $I_{1snp}$ is the $1s \to np$ photoexcitation energy. As a result, the probability of the $z(s,d)_{++}$– continuum-spectrum state being "ejected" by Coulomb's repulsion of the remaining $1s$–electron of the *K*-shell into the $ys_{++}$ – state of the continuous spectrum (Fig. 1 at $t = t_2$) is much higher than the probability of its ($1s$–electron) not being "noticed" (the "background" in Fig. 2).

## 4. Conclusion

Outside the framework of the dipole approximation, a non-relativistic version of the quantum theory of the effect of two-photon double ionization of the *K*-shell of a beryllium-like atomic ion has been constructed for the complete operator of the radiation transition between continuum-spectrum states. The isoelectron sequence of ions $Ti^{18+} \to Fe^{22+} \to Zn^{26+}$ has been studied. As the main result, in the high-energy region of absorbed X-ray photons $2\hbar\omega \in$ (12.5; 28) keV, a *gigantic* non-dipole effect was established in the calculation of generalized cross-sections from the leading $ss(^1S_0)$-, $sd(^1D_2)$–channels of the final states of double ionizaion of the *K*–shell of the ion. Taking into account correlation and relativistic effects is the subject of future development of the theory. The results of successful experiments on the observation of two-photon ionization of atoms, molecules, and solids (see [19 – 21] and references therein) suggest that the absolute values of the generalized cross-sections in Fig. 2 are quite measurable in the modern XFEL - experiment.

**Table 1.** The total width of the decay $1s$ – vacancies ($\Gamma_{1s}$; obtained from the theoretical data of the work [18]) and the energies of the thresholds of single ($I_{1s}$) and double ($I(1s^{-2})$) ionization of the *K*-shell (relativistic calculation of this work) of the ions under study.

| Parameters | $Ti^{18+}$ | $Fe^{22+}$ | $Zn^{26+}$ |
|---|---|---|---|
| $\Gamma_{1s}$, eV | 0.0850 | 0.0909 | 0.0967 |
| $I_{1s}$, keV | 6.0211 | 8.5571 | 11.5509 |
| $I(1s^{-2})/2$, keV | 6.2067 | 8.7791 | 11.8101 |

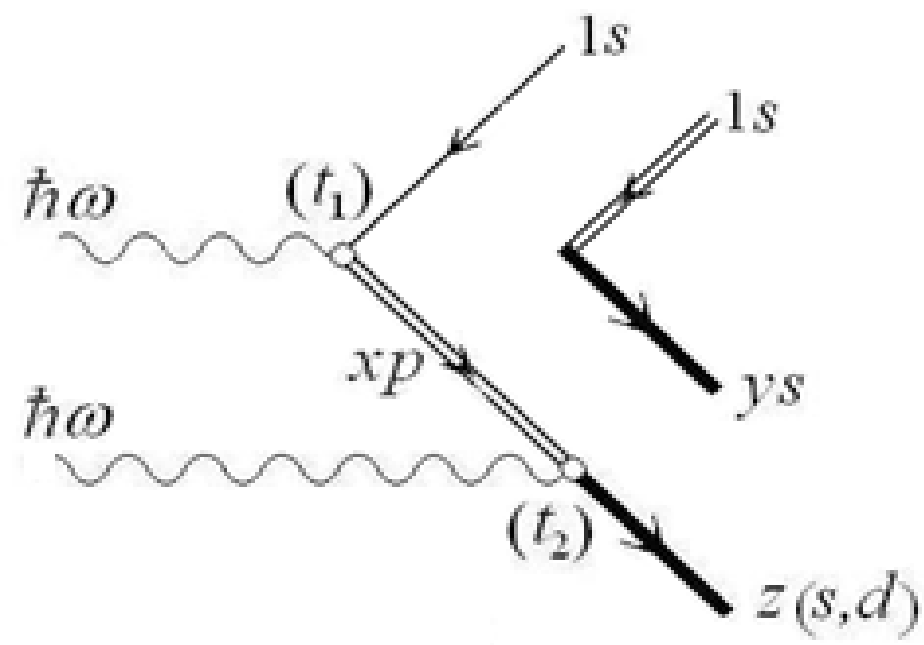

**Figure 1.** Probability amplitude of two-photon double ionization of the *K*-shell of a beryllium-like atomic ion over channels (1) in the representation of (non-relativistic) Feynman diagrams. The arrow to the left is a vacancy, the arrow to the right is an electron. A double (wide black) line is a state obtained in the Hartree-Fock field of one (two) $1s$-vacancies. The connection of the double and wide black lines corresponds to the overlap integral $\langle 1s \mid ys \rangle$. The light circle is the vertex of the radiation transition. The direction of time is from left to right ($t_1 < t_2$). $\hbar\omega$ is the energy of the absorbed photon.

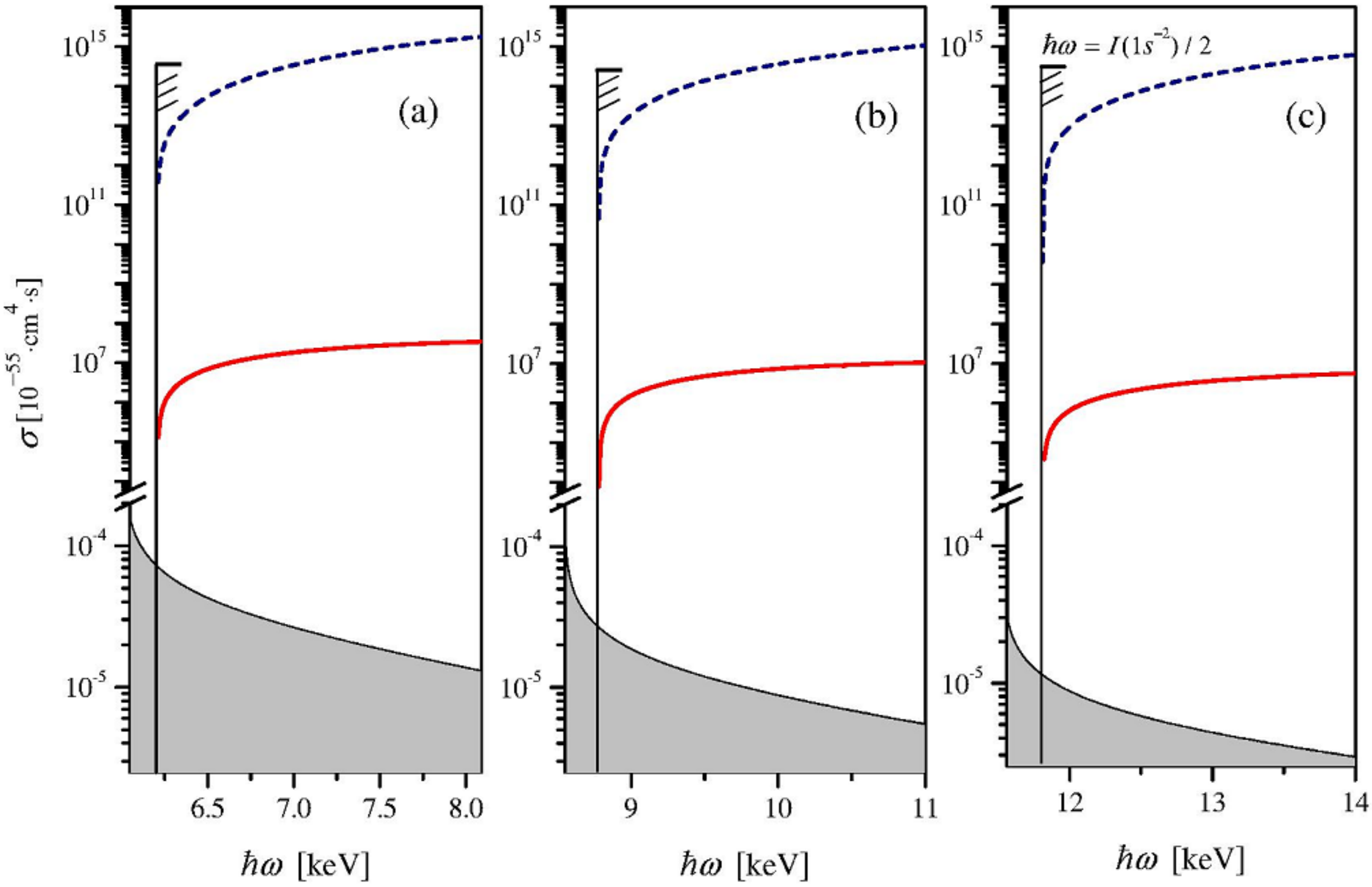

**Figure 2.** Generalized cross-sections of two-photon double ionization of the *K*-shell of beryllium-like atomic ions $Ti^{18+}$ (a), $Fe^{22+}$ (b) and $Zn^{26+}$(c) ($\sigma = \sigma^{(ss)} + \sigma^{(sd)}$): the dotted (solid) curve is the dipole (non-dipole) approximation for the radiation transition operator. The generalized cross-sections of the two-photon single ionization of the work [12] for the ions under study at the energy of the absorbed photon $\hbar\omega \in (6; 14)$ keV are given as the "background".